\documentclass[manuscript,screen,sigconf]{acmart}

\AtBeginDocument{%
  \providecommand\BibTeX{{%
    \normalfont B\kern-0.5em{\scshape i\kern-0.25em b}\kern-0.8em\TeX}}}

\acmConference{Under review}{XX}{XX}
\setcopyright{acmcopyright}
\copyrightyear{2023}
\acmYear{2023}
\acmDOI{XXXXXXX.XXXXXXX}

\acmJournal{JACM}
\acmVolume{37}
\acmNumber{4}
\acmArticle{111}
\acmMonth{8}

\acmPrice{15.00}
\acmISBN{978-1-4503-XXXX-X/18/06}

\usepackage{cleveref}
\usepackage{breakurl}
\begin{document}

\title[Reclaiming the Digital Commons: A Public Data Trust for Training Data]{Reclaiming the Digital Commons: \\A Public Data Trust for Training Data}
\author{Alan Chan}
\authornote{Corresponding author.}
\email{alan.chan@mila.quebec}
\affiliation{%
  \institution{Mila, Université de Montréal}
  \country{}
}

\author{Herbie Bradley}
\email{hb574@cam.ac.uk}
\affiliation{
  \institution{EleutherAI}
  \country{}
}
\affiliation{%
  \institution{University of Cambridge}
  \country{}
}

\author{Nitarshan Rajkumar}
\email{nr500@cam.ac.uk}
\affiliation{%
  \institution{University of Cambridge}
  \country{}
}

\renewcommand{\shortauthors}{Chan, Bradley, and Rajkumar}

\newcommand\todo[1]{\textcolor{red}{#1}}

\begin{abstract}
Democratization of AI means not only that people can freely use AI, but also that people can collectively decide how AI is to be used.
In particular, collective decision-making power is required to redress the negative externalities from the development of increasingly advanced AI systems, including degradation of the digital commons and unemployment from automation.
The rapid pace of AI development and deployment currently leaves little room for this power.
Monopolized in the hands of private corporations, the development of the most capable foundation models has proceeded largely without public input.
There is currently no implemented mechanism for ensuring that the economic value generated by such models is redistributed to account for their negative externalities.
The citizens that have generated the data necessary to train models do not have input on how their data are to be used.
In this work, we propose that a public data trust assert control over training data for foundation models.
In particular, this trust should scrape the internet as a digital commons, to license to commercial model developers for a percentage cut of revenues from deployment.
First, we argue in detail for the existence of such a trust. We also discuss feasibility and potential risks. Second, we detail a number of ways for a data trust to incentivize model developers to use training data only from the trust. We propose a mix of verification mechanisms, potential regulatory action, and positive incentives.
We conclude by highlighting other potential benefits of our proposed data trust and connecting our work to ongoing efforts in data and compute governance.
\end{abstract}





\maketitle

\section{Introduction}
Private companies dominate the development of the most capable AI systems \citep{giattino_artificial_2022}. The staggering amounts of compute involved \citep{sevilla_compute_2022,giattino_artificial_2022} mean that large tech companies or those backed by massive amounts of venture capital have disproportionate power in guiding the direction of technological progress. Recent attempts to democratize AI development and open up the study of large models have met with some success \citep{gao_pile_2020, black_gpt-neox-20b_2022, schuhmann_laion-5b_2022}, yet still suffer from core limitations. From a resource perspective, it remains difficult for academic or non-profit collaborations to match the financial weight of the private sector. From a philosophical perspective, democratization of AI is not solely about the free deployment of AI without regard for social consequence. Rather, we hold as \citet{shevlane_structured_2022} does that democratization also means collective decision-making power over how AI is to be developed and deployed. Narrow democratization could frustrate the broad democratic ideal; unstructured access to AI systems could hinder societies from restricting certain uses they deem undesirable.

Collective decision-making power over AI is deficient in two key respects. 
First, data creators cannot prevent AI developers from using their data. Opt-out mechanisms are lacking and the training datasets of many of the largest models are private. 
Second, there is no implemented mechanism to ensure that the profits of AI development and deployment are distributed widely, particularly as a way to redress negative externalities. 
Even if an individual were to threaten to withhold their data from a model developer, they would have effectively no bargaining power since a few data points likely make no significant difference in the final performance of a model. 

We focus on the large training datasets scraped from the \textbf{digital commons}---the collective intellectual and cultural contributions of humanity that are in digital form---and also on bespoke crowdworker data 
as a point of intervention for redressing the power imbalance between model developers and human data creators. 
The digital commons is the product of humanity's cumulative efforts, yet in AI development the fruits of the commons are captured by the few. Redistribution is requisite from the point of view of justice. Redistribution is also requisite from the point of view of pragmatism, for if human contributors to the digital commons are not supported in their work or resent its perceived theft, the commons itself could decay \citep{huang_generative_2023}.

To address the imbalance of power, we propose the creation of a public data trust. We intend this data trust to be national and located in a jurisdiction with a high concentration of AI development, such as the US or the UK. Our data trust would gate access to the most important data for model training: pre-training data from the internet and human feedback data from annotators. Our gating is meant to apply primarily to commercial AI developers. We focus our attention on general-purpose AI systems such as foundation models, given their likely role as important components of future AI systems and their increasingly wide adoption. 
Our contributions are as follows.
\begin{enumerate}
    \item We argue for the creation of a public data trust to hold training data, so as to address the private concentration of power in AI development and safeguard the digital commons. 
    \item We describe how the data trust could use its bargaining power to address the negative externalities of AI deployment, including setting up a digital commons fund financed by a royalty on model revenues.
    \item We propose how the data trust could obtain training data.
    \item We provide a detailed plan for how the data trust could verify that model developers who have agreed to the data trust regime have only used the trust's model in training their models.
    \item We discuss various mechanisms for incentivizing model developers to comply with the data trust regime.
    \item We advance other potential benefits of our data trust, including supporting the generation of training data as a public good.
\end{enumerate}

\section{The Case for a Data Trust}
We argue here for a national, public data trust to hold training data. An outline of our case is as follows.
\begin{enumerate}
    \item AI development heavily depends upon the \textbf{digital commons}: the collective intellectual and cultural contributions of humanity that are in digital form.
    \item AI development is extremely concentrated in the private sector. Those who contribute to the digital commons, including the general public and sector-specific individual such as artists, have little decision-making power over the development of deployment of AI compared to the AI developers.
    \item AI deployment results in negative externalities to the public; there are currently no effective mechanisms to address these negative externalities.
    \item No existing alternative effectively addresses the power imbalance.
    \item A data trust that gated training data access to the digital commons would help to correct the power imbalance so as to redress negative externalities.
\end{enumerate}

\subsection{The Digital Commons}
The \textbf{digital commons} \citep{dulong_de_rosnay_digital_2020,huang_generative_2023} constitutes the collective intellectual and cultural contributions of humanity in digital form. More specifically, the digital commons encompasses items like artistic work, scientific papers, knowledge bases, and software. Examples of resources that are a part of the digital commons include arXiv, Wikipedia, Reddit, online news sites, and Project Gutenberg. 

The digital commons is crucial for democracy, material well-being, and cultural enrichment. First, the success of democracy depends upon an informed public \citep{the_consilience_project_democracy_2021}. Absent an accurate understanding of the state of the world, the public is less able to engage in productive deliberation and to select representatives to act in their interest. Knowledge resources in the digital commons can contribute to this public understanding. For example, Wikipedia has been a surprisingly rich source of information, comparable to academically authored encyclopedias in both breadth and reliability \citep{mesgari_sum_2015}. Second, knowledge and tools in the digital commons contribute to material well-being. For instance, \citet{ghosh_economic_2007,directorate-general_for_communications_networks_impact_2021} characterize the large positive impact of open-source software on the economy of the EU. Third, the digital commons provides a source of intangible cultural enrichment. For example, on Project Gutenberg one can access over 60 000 works of intellectual and cultural significance, from the Federalist Papers to the Analects of Confucius. The role of the digital commons in these critical functions underscores the importance of safeguarding it.

\subsection{Concentration of Power}
The wealth of high-quality information in the digital commons is a prime source of training data for modern AI systems. Empirical scaling laws about the relationship between the quantity of data, compute, and model parameters \citep{kaplan_scaling_2020,hoffmann_empirical_2022} have motivated the use of ever larger amounts of data from the digital commons to train so-called ``foundation models'' \citep{bommasani_opportunities_2022, openai_gpt-4_2023}. Such models as GPT-3 \citep{brown_language_2020} are so named because they are increasingly general-purpose and seem likely to be deployed in a variety of scenarios \citep{bommasani_opportunities_2022}.

Given the enormous quantities of data and computation involved, private companies have a quasi-monopoly on the development of the largest---and by virtue of scaling laws likely the most capable---foundation models \citep{giattino_artificial_2022,ganguli_predictability_2022}. To obtain training data, private companies scrape the internet to obtain large datasets and hire crowdworkers to generate bespoke data. At no point is there an opportunity for the public to exercise decision-making power. Especially given the significant risks of AI development \citep{dafoe_ai_2018,bommasani_opportunities_2022,ganguli_predictability_2022}, private power to shape the trajectory of AI is in tension with public interests \citep{zuger_ai_2022}. Despite the proliferation of AI ethics standards in recent years \citep{jobin_global_2019}, ethical guidelines are no substitute for addressing the structural factors underlying the concentration of power in the private sector.

\subsection{Negative Externalities}\label{sec:neg-ext}
While concentration of power is itself suspect, the power of private AI developers contributes to tangible harms as well. Although the digital commons is the collective output of humanity, private organizations who train models on the digital commons stand to capture a large share of the profits while externalising the harms.


\subsubsection{Decay of the Digital Commons}
Given the political, social, and economic functions of the digital commons, its maintenance is paramount. While the increasing use of foundation models like ChatGPT and Midjourney can contribute to the digital commons by facilitating modes of artistic expression, they also threaten its degradation \citep{huang_generative_2023}. 

First, the widely available ability to generate content at scale threatens the quality of information in the digital commons. As language models (LMs) become more capable and access to them becomes cheaper\footnote{Access to the ChatGPT API as of 13 March 2023 is at \$0.002 USD / 1K tokens, which is about \$2 USD for 750K words.}, the scope and impact of misuse could increase. Politically motivated groups could use LMs to facilitate influence operations \citep{goldstein_generative_2023}. 
Even when used with the best of intentions, LMs still generate falsehoods that may be difficult to detect \citep{lin_truthfulqa_2021,ji_survey_2022}. Depending on how detection abilities scale with ease of generation, it may become more difficult to filter through online content for high-quality contributions. The problem of filtering is not only technical: even if capable tools exist for detecting low-quality contributions, we still need incentives in place for moderators to use those tools. The recent history of social media moderation shows that profit motives may override the importance of a high-quality public forum \citep{wetsman_facebooks_2021,perrigo_how_2021,wells_facebook_2021}. 

Second, a prevailing business model for foundation models may disincentivize contributions to the digital commons. This business model involves customers paying AI developers, such as OpenAI, for query access to their models. The developers capture the economic value of this transaction. Yet, since model developers externalize the costs of generating digital commons data, part of this economic value is rent, especially as private model developers are those best able to make use of large amounts of digital commons data to train models. Fees from using the text-to-image system DALL-E 2 go to OpenAI, not to the artists of the digital commons whose work was instrumental in the creation of such image models. People who otherwise might have hired artists might instead use DALL-E 2 for its lower cost.\footnote{As of 14 March 2023, users of DALL-E 2 receive 15 free generations every month and can purchase additional generations at a rate of \$15 USD per 115 generations.} When individuals use LMs as substitutes for search \citep{nakano_webgpt_2022}, they can obtain immediate answers which obviates visiting web pages. A decrease in ad revenue could negatively impact the sustainability of major sites in the digital commons like Stack Exchange. Using LMs as language assistants could also reduce the quantity of contributions on quality discussion forums like Reddit's \textit{r/AskHistorians} subreddit. 


\subsubsection{Unemployment}
Foundation models are not only able to generate text and images, but are increasingly capable of acting in the digital world. We are building language models that can code \citep{chen_evaluating_2021} and use arbitrary software tools \citep{schick_toolformer_2023}. In addition to the safety of such systems \citep{chan_harms_2023}, a key concern is the negative effects of widespread unemployment if these systems increasingly substitute for human labour. 

We are beginning to see these effects unfold. Companies use the work of artists from the digital commons to build text-to-image models, whose subsequent deployment deprives artists of the ability to make ends meet. The negative externality of unemployment exists even when training datasets are collected through crowdworkers and not the digital commons. Data from human programmers are used to improve coding models which threaten to substitute for the same programmers \citep{albergotti_openai_2023}.

The risk of mass unemployment is non-trivial given economic incentives to develop and deploy increasingly capable AI systems that could substitute for human labour at lower costs. \citet{korinek_preparing_2022} disarm common objections to the idea that machine labour could replace human labour in large portions of economic production. One objection extrapolates from the history of automation since the industrial revolution to claim that humans will just move to new jobs created in the wake of AI deployment. Yet the creation of such jobs in the past depended upon new demand for human cognitive labour. If AI development is to automate increasingly large amounts of cognitive labour, the future role of human labour is unclear. 
Despite any uncertainty over the precise shape of future employment, having mechanisms in place to address unemployment as a negative externality does not presume that everybody will be unemployed. Ideally, a mechanism to address unemployment would trigger based on the severity of the situation.

We emphasize that we are not arguing against the application of foundation models to increase productivity, improve well-being, and reduce the need for repetitive and unfulfilling labour. Rather, we are concerned about the distribution of the benefits and burdens of the AI development.


\subsection{A Data Trust to Control the Data Bottleneck}
There is currently no effective mechanism to address the power imbalance between private AI developers and the public, nor is there any mechanism to redress the negative externalities of AI development. Given the importance of training data, we propose that a public data trust should gate access to training data, both from the digital commons and crowdworkers. If the data trust is able to accomplish this task, it would hold significant leverage over private model developers. In effect, training runs of the most capable models would be severely hindered without access to data from the digital commons. Our focus here is on regulating commercial model development, rather than research use. 

A \textbf{data trust} is a legal vehicle for the collective management of data \citep{delacroix_bottom-up_2019}. In a \textbf{trust}, a board of trustees manages an asset on behalf of trustors, such as money, land, or buildings. Trustees typically have a fiduciary obligation to act only in the interests of the trustors. A data trust for training data would be composed of a board of trustees to manage collected training data on behalf of the public.

The data trust should be public because its decisions should reflect the public interest. The trustee board should be constructed so as to represent a diverse array of societal perspectives. Mechanisms such as regular reports to the legislature should be in place to hold the data trust accountable to the public. As our focus here is on the functions of a trust, we defer further details about the governance structure of the trust to future implementation.

The data trust should be national so as to have the authority to carry out its functions. 
In brief, the functions are as follows, with details deferred to \Cref{sec:obtaining-data,sec:verifying-compliance,sec:incentives}.
\begin{enumerate}
    \item Collect training data by scraping the internet and entrusting national user data.
    \item Implement a verification system to check that model developers who claim to be using the trust's data are only using the trust's data.
    \item Incentivize model developers to go to the data trust for data instead of scraping their own.
    \item Negotiate the terms of data usage with model developers, including royalties on a portion of revenue to go to national funds to support the digital commons. 
\end{enumerate}
Assuming that the data trust can accomplish (1) to (3)---the analysis of which we leave to later sections---the key part of our data trust proposal is (4). A data trust would negotiate the terms of training data usage with the public interest in mind, accounting for the negative externalities we raised in \Cref{sec:neg-ext}. For the rest of this section, we will assume that the trust has the necessary bargaining power to negotiate terms of data usage with model developers.

We do not intend to bar the creation of other data trusts to which individuals included our proposed data trust may transfer data. Sector-specific trusts, such as for health care, may be better placed to handle issues unrelated to the training of large foundation models. Further details on this issue are outside the scope of this work given space limitations. We will use \textit{the} trust henceforth.

\subsection{Addressing Threats to the Information Quality of the Digital Commons}

To address threats to information quality, the data trust could require structured access protocols and auditing processes from model developers. The AI community has experimented with a wide variety of access protocols in recent years \citep{solaiman_gradient_2023}. More structured protocols \citep{shevlane_structured_2022}, like only providing rate-limited API access, could make the generation of low-quality content at scale more difficult.  
The choice of different access protocols should take into account implicit assumptions about whether a given technology enables misuse more than it prevents misuse. \citet{shevlane_offense-defense_2020} argue that such conversations implicitly assume analogies to a particular field, such as software security, which may not capture the unique characteristics of AI development. 

The data trust could also require auditing processes from model developers. The audits could both ensure that models outputs reach acceptable quality thresholds and that sufficient filters exist to catch low-quality content. 
Both internal \citep{raji_closing_2020} and external \citep{raji_outsider_2022} audits on a regular basis would be helpful. Indeed, auditing is already a part of some proposed regulations on AI, such as the EU AI act \citep{noauthor_proposal_2021}. 


\subsection{Funds for the Digital Commons}
For both the problems of weakened incentives to contribute to the digital commons and unemployment, the key issue is that commercial model developers externalize the costs of the generating data in the digital commons. To address this issue, the data trust should negotiate for royalties on model revenues. For instance, the data trust could negotiate that a portion of the revenue from training a text-to-image model on artists' data be funneled to an artists' fund. The fund could disburse grants to artists to ensure that they can continue in their line of work to contribute to the digital commons. Such funds already exist in multiple jurisdictions. For example, the Copyright Board of Canada applies a levy to cassette and CD sales that is redistributed to Canadian artists \citep{noauthor_private_2023}. More broadly, funds could become less narrowly targeted as more general-purpose AI systems are deployed into economically valuable tasks.

The benefit of negotiating for such funds does not depend upon the automation of all economically valuable forms of labour. Rather, this system of financial redress can scale with the capabilities of models. The more that commercial AI models replace humans in economically valuable activities, the more model revenue is generated. Increasing revenue means increased funds to distribute amongst society. Moreover, such a fund could be implemented immediately as companies are already generating considerable revenue from model deployment, in contrast to a windfall tax \citep{okeefe_windfall_2020} which could only be implemented in the event of the deployment of a AI system with transformative economic impact.

\section{Potential Problems}
We analyze some reasons why a data trust might be ineffective at addressing the power imbalance in AI development.

\subsection{Political Will}
Because of the many activities our trust will have to undertake, the establishment of a data trust with enough power to execute its functions would likely require a substantial amount of political will. Yet, such will might already exist. Public entities are increasingly looking to regulate the development and deployment of AI systems \citep{noauthor_proposal_2021,noauthor_ai_2022,noauthor_establishing_2022}. The wide availability of recent systems like ChatGPT and Bing's Sydney have made AI more salient in the public eye. The ongoing lawsuit against Stability AI for using millions of photos from artists \citep{brittain_getty_2023} has brought to the fore ideas around redressing the negative externalities of AI development. 

\subsection{Model-Generated Training Data}
Although humans currently are responsible for generating most training data, recent advances in model-generated training data could threaten the centrality of human-sourced data \citep{bai_constitutional_2022,taori_stanford_2023}. \citet{bai_constitutional_2022} find that the use of model-generated feedback data for reinforcement learning fine-tuning provides a Pareto improvement in harmlessness and helpfulness compared to using only human-generated feedback data. Moreover, \citet{wu_insights_2022} find that synthetic pre-training datasets can provide a significant portion of the benefits of human-sourced pre-training sets. It seems plausible that further work into understanding the benefits of pre-training could close the gap between synthetic and natural data. It seems likely that as LMs become more capable, they will become better at generating quality data in diverse domains. 

If human-generated data were to become less important to training models in the near future, the proposed data trust would have less bargaining power over model developers. 
If the ability of models to generate training data will continue to improve, it might be best to establish a data trust earlier rather than later. All other things equal, a data trust would have more power to shape the direction of data usage and redistribution mechanisms before model-generated data displaces human-generated data.


\subsection{Corporate Capture}
The private sector is extremely well-funded. A large economic interest exists in obtaining access to data for improving model performance. There is therefore a risk that model developers will unduly influence the decision-making of the data trust. Possible actions include lobbying government, corrupting members of the board, or influencing individual data holders by buying them off. Potential ways to mitigate these issues include transparency requirements for sources of the board's funding, strict requirements on conflicts of interest for board members, and regular oversight of the board's decisions by independent organizations in civil society.

\subsection{Government Capture}
A data trust should be insulated enough from government to make decisions based truly upon the public interest, rather than upon ephemeral political winds. Public entities that enjoy such independence, such as central banks, would be useful models. 

Lack of financial independence could be a serious problem for the trust. Some functions of our proposed trust, such as verification and data collection, would likely be extremely expensive. Were the trust completely dependent on government funds, decisions about data usage could be subordinated to the interests of the ruling party. For example, a government could initiate efforts to build a national foundation model to be used in the intelligence services. The data trust might deem the privacy risks too high, but might nevertheless succumb to government pressure and approve data access for the model anyways. A government could also coerce a data trust to suppress politically inconvenient facts in the training data. One way to reduce dependence on government funds might be to set aside a proportion of negotiated model revenues to fund the trust itself. 

\section{Obtaining Data for the Trust}\label{sec:obtaining-data}
Having made the case for a data trust, we now go into implementation details. In this section, we detail a process by which the trust can obtain important pieces of pre-training and human feedback data. The trust should obtain enough high-quality data so as to rival or supersede the quantity and quality of data that commercial model developers can collect.

\subsection{Sources of Training Data}\label{sec:data-sources}
To understand how a data trust would operate, we review the key sources of training data that data trusts should target for control. 

\subsubsection{Pre-Training Data}
\textbf{Pre-training} is the process of performing self-supervised learning with a foundation model on a large corpus of text. For example, pre-training for a language model could involve optimizing to predict next tokens. Pre-training on large corpora of data has been responsible for many of the massive improvements in AI capabilities in the past 5 years \citep{wei_emergent_2022, brown_language_2020, openai_gpt-4_2023}. For large language models, pre-training dataset sizes can run into the trillions of tokens and over 5 TB of pure text \citep{hoffmann_empirical_2022}, while for image models they can be as large as 4 billion images \citep{dehghani_scaling_2023}. Given empirical scaling laws that provide predictable relationships between compute, data, model size, and performance \citep{hoffmann_empirical_2022}, training on increasing amounts of data is currently the clearest path to improving model capabilities. Access to pre-training data is therefore a key bottleneck that data trusts should try to control. 


Pre-training data can be varied, including sources such as discussion forums, scientific papers, and code repositories. Much of this data is freely available on the internet. Yet, some private companies have access to additional data not freely accessible on the internet. 
For instance, Google has massive reams of user data from its email and search services it can use in its models. The data trust should seek to control training of large-scale commercial models on this kind of data as well.

\subsubsection{Human Feedback Data}
\textbf{Human feedback data} refers to any type of signal that indicates human preferences over the data distribution. For example, one type of human feedback is in the form of high quality human examples---when training a model to summarize articles, developers might obtain human-written reference summaries to fine-tune their model on so that the model output more closely aligns with human preferred summaries \citep{stiennon_learning_2020}.

Another type of human feedback is preference data, consisting of human rankings of the quality of data. These preferences can be used to train a reward model, which in turn can be used to fine-tune a foundation model, in a process known as reinforcement-learning from human feedback (RLHF) \citep{christiano_deep_2017,stiennon_learning_2020}. High-quality human preference data has proven to be extremely effective for fine-tuning large language models to be more helpful and harmless \citep{bai_training_2022}. 

Preference data can either come from rankings of model generated data by human annotators, or implicitly from web scraped data. In the former case, model developers will typically pay a specialised AI data collection vendor such as Scale AI or Surge AI, or alternatively hire crowdworkers themselves via platforms such as Amazon Mechanical Turk. In the latter case, developers may scrape public internet forums, such as Reddit, to obtain implicit preferences from metadata such as votes or likes \citep{ethayarajh_stanford_2023}.




\subsection{Scraping Data}
The data trust should scrape the internet to construct its own large-scale pre-training datasets. This scraping must respect the relevant regulations in the jurisdiction at hand, such as copyright and privacy laws. To perform this scraping, the data trust could partner with organizations that have relevant expertise, such as EleutherAI \citep{gao_pile_2020}. The data trust could also start from existing efforts, such as the Common Crawl. We emphasize that the process of scraping data should be a continual, iterative process given the continual growth in the amount of internet data \citep{villalobos_will_2022}.

The data trust should curate and document the collected data in detail, following best practices \citep{gebru_datasheets_2021,hutchinson_towards_2021,mitchell_measuring_2023}. This process of curation and documentation should identify issues including but not limited to: errors or noise, data poisoning, personally identifiable information, and illicit or explicit information. The choice of data to exclude from a pre-training set can be difficult. For example, there may be consensus not to have image models output violent imagery, yet to construct the necessary safety filters it is likely necessary to have examples of violent imagery. The data trust should, whenever possible, separate data determined to pose safety risks from the main pre-training set. Since the act of doing so is inherently value-laden, the trust should carry out this process through or under the supervision of a diverse panel of experts across disciplines, with explicit representation of voices from marginalized communities. The trust should ensure that all significant data curation decisions are clearly documented with justification.

\subsection{Obtaining Data that Cannot be Scraped}

\subsubsection{Restrictive or Non-Existent Licenses}
Some publicly available data reside on large community sites, such as DeviantArt or Reddit's \textit{r/art} subreddit. Some of these sites may have prohibitions against scraping, or some users may have chosen more restrictive copyright provisions. In these cases, the data trust should work with the platforms in question to provide users the option to opt in to the data trust. Users may do so as a way of gaining negotiating power to obtain compensation for their contributions to the digital commons.

\subsubsection{Obtaining User Data}
Beyond community platforms, large tech companies such as Google, Meta, and Twitter hold vast amounts of user data that would be useful for training foundation models, if the companies themselves do not already use them or license them out. Some of these platforms hold large market positions, such as Google for email \citep{noauthor_gmail_2022} and Meta for social media \citep{ortiz-ospina_rise_2019}. Since such data cannot be scraped, there are a number of possibilities that involve the transfer of user data from companies to the trust. 

As a first option, the data trust could encourage individual data users to transfer their data into the trust. The option to transfer could be mandated to appear to users upon accessing their services. The data trust could engage in a public outreach campaign to encourage such transfer, which might meet with some success given popular suspicion of big tech companies \citep{kelly_americans_2021,fischer_tech_2022} Although this method would be the least forceful, it might suffer from low uptake given user inertia, a lack of interest, or ignorance about data governance \citep{delacroix_bottom-up_2019}.

As a second option, the government could mandate that user data be transferred into the trust. 
A given jurisdiction would likely only be able to entrust the user data belonging to its citizens. Nevertheless, there might be ample data anyways. The population of the United States is more than 300 million, while the population of the EU is more than 400 million \citep{roser_world_2013} While mandating data entrustment may appear radical, it is only because we are used to the status quo. Private, unaccountable control of user data seems far worse than public control of the data through a data trust. Especially since terms of service can be so long and difficult to understand that many skip them entirely \citep{obar_biggest_2020}, it is likely that many users did not provide meaningful consent for platforms to hold their data. 


\subsubsection{Obtaining Human Feedback Data}
To obtain human feedback data, data trusts could work with both crowdworker collectives \citep{irani_turkopticon_2013} and crowdsourcing platforms like Surge and Upwork to include human feedback data from crowdworkers in the trust. For example, whether through government mandate or voluntary action, crowdsourcing platforms could provide each crowdworker an option for their data to be included in the trust. Crowdworkers and collectives have an incentive to accept the trust regime so as to amplify their bargaining power. Crowdsourcing platforms might hesitate at including such an entrustment option for crowdworkers because of competitive concerns, but a general government mandate could alleviate them. 

\section{Verifying Compliance}\label{sec:verifying-compliance}
To obtain leverage, the data trust needs to ensure that model developers only use data from the trust. We consider it infeasible to ban scraping outright. Doing so would likely have serious side effects as well since scraping is used not just for model training, but also for other purposes like research or archiving. 

Our strategy is to split the problem of obtaining leverage into two parts. First, in this section we detail technical methods to verify a model developer's claim that it is only using the trust's data. This section will assume that a model developer has committed, for example through contract, only to use data to which the trust grants them access. The question is how to enforce such a commitment.
Our technical methods involve the following steps.
\begin{enumerate}
    \item Anybody who obtains data from the data trust actually trains the model with the trust's data.
    \item The data trust's dataset is the only dataset used to train the model. 
    \item When the model developer deploys the model, the deployed model is the same as the trained model that the data trust verified. 
\end{enumerate}
Second, in \Cref{sec:incentives} we explore a variety of options for incentivizing model developers to comply with the data trust regime.





\subsection{Verifying that the Trust's Dataset was Used}

Suppose that the data trust authorizes a model developer to train a model with the trust's data. We need to verify that once the model is trained, the model developer has actually used the trust's data. Our proposed method involves inserting digital signatures into training sets that the trust provides to model developers, based heavily on existing work in data poisoning attacks \citep{carlini_poisoning_2022,carlini_poisoning_2023}. 

\subsubsection{Inserting Digital Signatures}
In data poisoning \citep{tian_comprehensive_2022,cina_wild_2023}, an adversary modifies a training set so that a model trained on this set will return a chosen output given a specific input. For example, it is possible to modify just 0.01\% of an image-caption dataset to cause a model to output an arbitrarily chosen caption on a select image \citep{carlini_poisoning_2022}. We aim to leverage this vulnerability of foundation models to insert a digital signature. 

The data trust shall generate a set $Y := \{(x_i, y_i)\}_{i = 1}^n$ of input-key pairs, where $x_i$ is an input to the foundation model and $y_i$ is a secret key. We call each $(x_i, y_i)$ a \textbf{digital signature}. $Y$ is therefore a set of digital signatures. $Y$ should remain unknown to the model developer. Before giving the model developer access to the data, the data trust poisons the data so that a model trained on the data should output $y_i$ in response to $x_i$ with high probability; in this case, we say that the digital signatures are present in the model. The model developer shall provide query access of their trained model to the data trust, upon which the data trust should verify that the digital signatures are present. Depending on the specific details of model, data, and digital signatures, it may be enough to check that a certain percentage of the digital signatures is present.


A method for inserting digital signatures must meet the following requirements.
\begin{enumerate}
    \item It should be computationally difficult to detect which pieces of training data are the digital signatures.
    \item A model trained even for only one epoch on the poisoned data should output each digital signature with high probability. 
    \item A model not trained with the poisoned data should only output each digital signature with low probability.
    \item The insertion of data signatures should not negatively affect the trained model's performance in a significant way. 
\end{enumerate}

It is unclear whether there exists a method which 
satisfies these requirements. We detail some initial proposals for text and image models, either based on or inspired by the techniques in \citet{li_open-sourced_2020,carlini_poisoning_2022,carlini_poisoning_2023}. We mean these proposals as initial ideas to be tested and iterated upon. We also note existing work on data poisoning for RL models \citep{rakhsha_policy_2020,lu_adversarial_2022,gunn_adversarial_2022}, which may be useful for inserting a digital signature into human feedback data.

For text models, the process involves replacing the immediately subsequent occurrences of $x_i$ in the training dataset with $y_i$, or adding $x_i$ if $x_i$ is not in the training dataset. The process for image models is similar. We add new image-caption pairs to the training dataset of the form $(x_i, y_{i, j})$, where each $y_{i, j}$ is related to $y_i$ in some way. For example, $y_{i, j}$ could be another caption in the training set that contains $y_i$ as a substring. 

\subsubsection{Potential Issues with Digital Signatures}
Model developers could work around the data poisoning in a number of ways. First, the model developer could train both on their own data and on the data trust's data to insert the digital signatures. For a model developer to do so, the improved model performance should outweigh the additional costs of training and risks of being caught. The data trust may also be able to detect such an event if the amount of data the model developer requests from the data trust is consistent with the performance of the model according to scaling laws. 

Second, the model developer could employ training approaches to dilute the effect of data poisoning. \citet{geiping_what_2022} show that interweaving data poisoning into adversarial training can protect against data poisoning attacks with a mild performance penalty for the model. Since \citet{geiping_what_2022} target image classification, it remains to be seen how effective such defenses would be on language and text-to-image models. \citet{wallace_concealed_2021} show that early-stopping can provide a moderate defense against data poisoning in language models at the cost of some predictive accuracy. Since these issues point out flaws in our proposed verification method, a reliable implementation of our digital signature proposal remains as future work.

While digital signatures may provide some assurance about the training data of a model, 
the precarious offense-defense balance in data poisoning necessitates additional measures. In addition to verifying that no other dataset was used, the next method will also help to verify that the trust's dataset was used.


\subsection{Verifying that No Other Dataset was Used}\label{sec:pol}
We now need to verify that no other data was used to train the model. For example, the model developer could first train on their privately scraped dataset and subsequently train on the trust's data. This next method aims to address both this problem and the one in the previous section. Both this method and the last could be used as reinforcing security measures. 

We use the proof-of-learning (PoL) framework that \citet{jia_proof--learning_2021} propose. In the PoL framework, the data trust requests a \textbf{proof} from the model developer, consisting of an encrypted set of model checkpoints $\{ (W_i, I_i, A_i) \}_{i = 0}^T$, where $W_i$ are the weights, $I_i$ are the indices of the data used to obtain $W_i$, and $A_i$ is auxiliary information such as optimizer state. Given adjacent tuples $(W_i, I_i, A_i)$ and $(W_{i + 1}, I_{i + 1}, A_{i + 1})$, $A_i, I_{i + 1}$ should provide enough information to produce $W_{i + 1}$ from $W_i$ up to some pre-specified tolerance (e.g., because of hardware randomness). $W_0$ is the model initialization and $W_T$ is the final model. 

Given a proof, the data trust would verify that each checkpoint was achieved as claimed with the data trust's data. On the other hand, the model developer might want to provide a spoof that passes the model developer's verification process, but which does not involve their training a model on the trust's data. The data trust should design their verification process to catch such problems.

PoL consists of the following steps.
\begin{enumerate}
    \item Verify that $W_0$ is a random initialization with a statistical test. We would not want $W_0$ to be pre-trained on a private dataset.
    \item Select indices $i_k$ to verify.
    \item For each $i_k$, start from $W_{i_k}$ and use the $A_{i_k}, I_{k + 1}$ to train until the timestep associated with $W_{i_k + 1}$. Call this new weight $\tilde W_{i_k + 1}$.
    \item If $\tilde W_{i_k + 1}$ is sufficiently different from $W_{i_k + 1}$, reject the proof.
\end{enumerate}
Running the above process for all indices $i \in [T]$ just reproduces the training process. Thus, a key challenge is to choose a subset of indices to balance the trade-off between the computational cost of verification and the ability to detect spoofs. \citet{jia_proof--learning_2021} propose a heuristic of selecting the pairs of checkpoints which resulted in the largest weight updates, but there is as yet no method with a formal security guarantee \citep{jia_proof--learning_2021,fang_fundamental_2022,zhang_adversarial_2022}.

Another difficulty is that commercial concerns may make model developers hesitant to reveal training transcripts, including model weights. Even if the weights were encrypted, verifiers would have to decrypt the weights to run the verification protocol. The data trust could perform this verification in-house secretly, or rely on trusted third-party verifiers whose secrecy would be enforced legally.

\subsection{Verifying that the Deployed Model is the Trained Model}
The previous methods attempt to verify that the model developer indeed has trained a model only on the data that the trust has provided. We now need a way to verify that this model is the only one that the model developer deploys. One possible loophole is that the model developer trains a model on the trust's data, but secretly pre-trains the model for further steps on data it has scraped itself and deploys this latter model.

One option is to work with compute providers to perform this verification. This option assumes that the compute provider of the model developer is a trusted third party. If the PoL verification of \Cref{sec:pol} succeeds, the data trust could transfer a hash of the final model weights to the compute provider. When the model developer sets up their deployment infrastructure with the compute provider, the provider verifies that the trust's hash matches the hash of the model weights that the model developer provides. If not, the compute provider refuses to deploy the model and notifies the data trust, who initiates regulatory action.

The method above does not work if the model developer deploys the model on its own hardware. Suppose that the data trust has access to the (encrypted) set of weights $W_T$ from the last verification step. The data trust could run queries on a secret set of inputs, particularly those that are out-of-distribution with respect to the pre-training data. \citet{janus_anomalous_2023} provide some evidence that distinct models have different log-probability distributions on out-of-distribution inputs. 
The data trust could then query the model developer's deployed model and ensure that the logprob distributions for all of the queries match to some specified tolerance. 

One difficulty with checking queries is that model developers may add noise or watermarks to their deployed models to ensure that others cannot copy the model easily \citep{boenisch_systematic_2021,gu_watermarking_2023,kirchenbauer_watermark_2023}. In this case, it seems like asking the model developers for their noise and watermark methods would not be too onerous, especially if it allowed data trusts to ensure that the model developer is following its commitments.

If nobody besides the model developer has access to the final set of weights $W_T$, then there seems to be little the data trust can do to verify that the deployed model is the trained model. This gap is a limitation of our proposal.



\subsection{Additional Problems with Verification}
We identify some additional problems with the effectiveness of our verification regime. 

\subsubsection{Cost}
Performing all of our verification steps is likely to be an expensive endeavour. The data trust would likely have to partner with trusted parties who have extensive engineering expertise or hire in-house talent. Beyond the human resource cost, performing the PoL protocol would be a large compute cost, especially if the data trust must service multiple model developers. Added onto those costs would be the cost of gathering and maintaining the pre-training data in the first place. 


\subsubsection{Leakage of the Trust's Data}
We do not want model developers to leak training data that the trust has provided to them. Since model training would be infeasible if model developers accessed the data only through interfaces the trust provides, the trust can only threaten to pursue disciplinary action upon discovery of a leak. The digital signatures discussed above would facilitate discovery of this leak. Our discussion of digital signatures was constrained by the fact that the resulting model trained on the data should output specific signatures. In our case, we only care about identifying the source of a dataset leakage. The design space is thus more open here and we can take advantage of continuing work on dataset watermarking \citep{li_open-sourced_2020,li_untargeted_2022,tang_did_2023}. 

\subsubsection{Small Teams of Model Developers}
It is difficult to prevent individuals or small teams of model developers from scraping some internet data and training a model. Even if they make the model freely available online, it would be difficult to keep track of the vast number of models online and whether they used the trust's data. Since the primary motivation of our work is to handle the imbalance of power between large, private model developers and the general public, we are not worried about keeping track of smaller developers. 

\subsubsection{Open-Source Developers}
One potential loophole is if commercial model developers work with non-commercial or open-source researchers and developers to create models for them. The commercial model developer could scrape the data it wants and provide it to the non-commercial developer, for example an open-science non-profit like EleutherAI \citep{phang_eleutherai_2022}. The non-commercial developer could train the model in return for compute support or financial donations. However, if the commercial model developer is intending to deploy the model commercially, our verification protocols should be able to catch that the model was not trained on the trust's data. 

Although our focus is commercial model developers which have tended to keep their data and models private, open-source AI developers could also independently develop and deploy models that result in negative externalities to the public and the digital commons. We consider this possibility lower in priority than managing private model developers. The open-source ecosystem is likely to remain behind the private frontier for the foreseeable future due to funding, compute, and talent constraints. Even once a frontier model has been developed, ongoing inference costs to deploy the best quality models to millions of people---which dwarf training costs \citep{patel_inference_2023}---are an additional reason for private developers to remain the central concern.


\subsubsection{Updating Deployed Models}
Model developers may routinely update their deployed models in response to user feedback. For example, the ChatGPT interface lets users provide binary feedback on generated output. This function is likely commercially important to model developers. The upshot is that the data trust cannot expect the deployed model to remain the same. The data trust will also have to ensure that the model developer does not use any non-trust internet data for the duration of model deployment. Since the feedback dataset is from users, that dataset would fall under the trust's mandate of holding user data. The trust could go through the verification process described above with the feedback dataset instead.

\section{Incentives to Submit to the Data Trust Regime}\label{sec:incentives}
Up until now, we have discussed technical methods for verifying a model developer's claim that they have complied with the demands of the data trust. Now, we discuss what the data trust should do to incentivize the model develop to submit voluntarily to the data trust regime. 

\subsection{Regulation}
Regulation could stipulate that authorization from the data trust be necessary for training a model on internet-scraped pre-training data for commercial usage. Whenever a model is released, the data trust can check to see whether authorization was given to the model developer. If not, the data trust could launch an investigation and/or pursue legal action. If yes, the data trust could proceed with the verification mechanisms in \Cref{sec:verifying-compliance}.

Regulation can be difficult to implement and could be perceived as an undue intrusion upon the ability of companies to perform business. At the same time, some amount of regulation will likely be necessary given the large incentives to capture the economic value of AI deployment. 
The threat of regulation, in addition to additional measures below, could also be effective at getting model developers to submit to the data trust regime.

\subsection{Certification}
As an alternative to regulation, the data trust could provide certifications for companies that voluntarily agree only to use the trust's data and submit to the verification regime in \Cref{sec:verifying-compliance}. Such a certification would work similarly to how Fair Trade labels \citep{dragusanu_economics_2014} do. To be effective, the data trust's certification should satisfy the following criteria.
\begin{enumerate}
    \item Consumers can easily distinguish between model developers who have certification and those who do not.
    \item There are consumers that care about model developers having certification.
    \item The buying power of consumers who care about certification is enough to offset the increased cost of a model developer's complying with certification requirements.
\end{enumerate}
We now argue for why the data trust's certification could satisfy these criteria. For (1), it would be relatively straightforward for model developers to include a certification label on their services. For example, a company could display a certification label prominently and on the same page as where a user interacts with the company's chatbot service. Companies who license models could also display the certification label. For (2), it seems plausible that a large proportion of citizens are interested in certification, especially given the prominence of data privacy issues \citep{mccallum_meta_2022,touma_tiktok_2022,information_commissioners_office_ico_2022} and controversies over unfair compensation for data generation \citep{perrigo_exclusive_2023}. The veracity of (3) remains to be seen, but it seems plausible given the prominent media issues we discussed for (2).


\subsection{The Data Trust's Comparative Advantage}
There are also positive incentives for model developers to accept the data trust regime. Data collection tends to be an arduous, costly process. Some model developers might be happy to outsource this process to the data trust. Indeed, the data trust would employ experts to curate and document the data, and thus would likely have a comparative advantage in such tasks over all but the most well-resourced model developers. Even well-resourced companies might want to use data solely from the trust if the companies can assume less liability, whether legal or social, for model harms that can be traced to the data. 

Additionally, when scraping the open internet through resources like the Common Crawl corpus, the vast majority of data does not come with a license attached, and is therefore considered ``all rights reserved'' by default. The vast majority of internet-scraped data is in this form, and so projects attempting to scrape only openly-licensed content are restricted to only a small fraction of Common Crawl. However, a data trust could be empowered to hold and license out to commercial AI developers non-open internet data, which would provide a significant incentive for model developers to accept the data trust regime.

\section{Other Benefits}
We note here other benefits of our data trust regime which are not related to our main benefit of addressing power imbalances.

\subsection{Test Set Leakage}
Test set leakage is when the evaluation set contains parts of the training set. In the presence of test set leakage, the evaluation metrics of a model become biased. Since evaluation datasets are often public \citep{bowman_snli_2015,srivastava_beyond_2022,gao_framework_2021} and pre-training sets are often private, it is difficult for the public to verify that foundation model evaluations are unbiased. If the data trust were to gate access to pre-training data, it could ensure that the pre-training datasets not contain any evaluation data. In particular, the data trust could hold a separate category of particularly important evaluation data, including evaluations on toxicity \citep{gehman_realtoxicityprompts_2020}, truthfulness \citep{lin_truthfulqa_2021}, and power-seeking tendencies \citep{perez_discovering_2022}.


\subsection{Supporting Opt-Out Mechanisms for Privacy}
The EU's GDPR recognizes that data subjects have a right to the erasure of their personal data. Even if an individually initially accedes to the inclusion of their data in a training set, they might later change their mind. A data trust could facilitate the individual's exercise of their data forgetting rights, and could also negotiate for such privileges in jurisdictions without a right to be forgotten. 

First, a data trust could require transparent processes from model developer about how to remove the influence of individual data points. At the same time, the field of how to do so is still evolving \citep{nguyen_survey_2022}. 
Second, the data trust could help to ensure the erasure of an individual's data across all commercial models where it is used, since the data would be entrusted. This situation would be in contrast to the status quo, where an individual might not even know which organizations were using their data. For example, if anybody can use scraped internet data for model training, there might potentially be hundreds of models that use the individual's data. Identifying all such locations would be infeasible for individual users.

\subsection{Supporting the Generation of Public Goods}
In addition to collecting generic data for training foundation models, the data trust could also support the collection of data that would be public goods. As an example, we focus here on the safety of AI systems as a public good. 

\subsubsection{Safe AI Systems as a Public Good}
In this section, we broadly construe safe AI systems as those that are steerable \citep{bai_training_2022} and that inhibit clear misuse such as political violence. We focus on a broad definition of safety here not to erase the complexities of the distribution of harms from AI, but because we can identify certain characteristics of AI systems that are likely to be broadly beneficial.

The safety of AI systems is a public good. Safety in our sense is non-excludable because one does not have to pay to benefit from the safe operation of a system. Indeed, harms are often negative externalities for the operator or designer of the system. Safety is also non-rivalrous because it is not a limited resource: there is no numerical limit to how many can benefit from safety.

\subsubsection{Free-Riding}
Certain kinds of training data likely contribute significantly to the safety of AI systems. For example, human preference data to increase the harmlessness of models \citep{bai_training_2022} likely increases safety. Let us call such data \textbf{safety-enhancing}. Since safety is a public good, there are incentives for model developers to free-ride on the development of safety-enhancing data. Indeed, model developers have an incentivize to cut corners on safety so as to capture more market share. For example, Microsoft was the first major player to integrate a chatbot into its search engine, but the chatbot has acted in an aggressive and manipulative manner \citep{vincent_microsofts_2023}. Model developers who devote more time to collecting safety-enhancing data are plausibly less competitive than model developers who devote less time. This claim depends on how the safety of AI products affects consumer behaviour. More work needs to be done to study this uncertainty.
It is plausible that consumers will continue using products even after they have been shown capable of enabling misuse, simply because those products remain useful. 

Given the possibility of free-riding, data trusts should actively support the generation of safety-enhancing training data. Making such training data publicly available serves two purposes. First, such datasets are often extremely expensive to generate. Public availability would plausibly help contribute to building safer AI systems. Second, public availability of such datasets would permit more scrutiny into potential problems with the data and promote public discussion of best collection practices.


\subsubsection{Generating Public Goods}
The process of collecting safety-enhancing data can be split into identifying which data would be public goods and collecting the data. Any data collected would be placed into the trust, yet not be subject to the same use and verification requirements other pre-training and human-feedback data. Instead, the data would be public.

Identifying safety-enhancing data, and other data as public goods, would likely require ongoing consultation with diverse communities and experts across disciplines. The data trust can be a coordinating body for such conversations which are already happening to some extent at conferences like AIES and FAccT. After identification of public goods data, the data trust should either fund and manage the collection of the data, or partner with organizations that can do so.

\section{Related Work}
\subsection{Data Governance}
In recent years, a number of jurisdictions have introduced legislation enshrining various rights of data holders, including the EU's GDPR, Canada's PIPEDA, and California's CCPA. Part of the motivation of such legislation has been the increasingly apparent ways in which tech companies may misuse personal data \citep{mccallum_meta_2022,touma_tiktok_2022,information_commissioners_office_ico_2022}.

\citet{delacroix_bottom-up_2019} propose data trusts as legal vehicles to exercise data rights on behalf of data holders as fiduciaries. \citet{viljoen_relational_2021} critiques the idea that data rights are an individualist notion, arguing that data are often relational. Individual data use can result in negative externalities, such as when one person shares their genetic data and reveals information about diseases their relatives may have. Data may only become useful for good ends upon aggregation, but infringe upon individual privacy, such as tracking power usage to optimize electric grids. Addressing these concerns requires collective vehicles to govern data usage.

The implementation of data trusts has so far preliminary. Data trusts have been explored in areas such as health\footnote{\url{https://www.ukbiobank.ac.uk/}}, cities \citep{scassa_designing_2020}, and finance\footnote{\url{https://www.openbanking.org.uk/}}. To our knowledge, there are no existing initiatives to implement data trusts for training data, although several works allude to the possibility \citep{delacroix_democratising_2020,zygmuntowski_embedding_2021,huang_generative_2023}.

\citet{huang_generative_2023} is the closest work to ours. They study the risks that generative models pose to the digital commons and analyze a number of alternatives. Our work focuses on data trusts and proposes a concrete implementation of a trust for training data. Another highly related work is \citet{jernite_data_2022}, which provide a framework for the governance of language model data. They also propose a data stewardship organization to establish and formalize relationships between actors in data ecosystems, which may span across nationalities. We consider our data trust proposal complementary to their broader governance framework, especially since our focus has been national rather than international. 

\subsection{Data Quality}
In addition to broader questions around the use of data, substantial research has investigated the quality of the training sets of AI systems. A particularly salient question has been the degree to which training sets reflect negative characteristics of human societies, including inequality, toxicity, and violence, and to what extent such characteristics are passed onto models \citep{buolamwini_gender_2018,abid_persistent_2021,harris_exploring_2022,wolfe_american_2022,gehman_realtoxicityprompts_2020}. The fact that models do indeed reflect parts of their data has motivated the development of tools and frameworks for better data documentation and creation practices \citep{gebru_datasheets_2021,mitchell_measuring_2023,hutchinson_towards_2021,von_werra_evaluate_2022}, so as better to understand and mitigate harms.

\subsection{Compute Governance}

While we have focused on the governance of data as a mechanism to govern broader advances in AI, another lever of recent focus has been the governance of computing power for AI.
As the most capable AI systems make use of exponentially increasing amounts of compute, now doubling every 10 months at the frontier \citep{sevilla_compute_2022}, control of computing power could provide an effective means of controlling AI system development and usage as well as broader progress in the field \citep{hwang_computational_2018}

Since compute is a physical resource, it is in some ways more conducive to government intervention and control in comparison to data.
At present there is little such regulatory intervention however, and furthermore there is a significant lack of even basic measurement or monitoring capability of how this resource is used for model training \citep{whittlestone_why_2021,noauthor_blueprint_2023}.
The physical nature of compute also has drawbacks in relation to an approach focusing on data -- stronger government intervention on compute, such as through National AI Research Clouds \citep{ho_building_2021}, would cost on the order of hundreds of millions or billions of dollars \citep{noauthor_strengthening_2023}.

\section{Conclusion}
Through data, the construction of today's most advanced AI systems depends upon the cumulative intellectual and cultural contributions of humanity. Yet, the public holds relatively little power over the conditions of AI deployment. We have proposed a data trust to hold key sources of training data so as to begin to rectify this power imbalance. Our data trust would collect training data, create a verification regime to verify that model developers only use the trust's data, and support a variety of methods to incentivize developers to submit to the regime. 

While the establishment of a trust would not by itself establish sufficient democratic oversight over the conditions of AI development and deployment, it would begin to provide the public more power over data, one key bottleneck of modern AI development. So as to ensure broad distribution of the fruits of AI progress, future work should aim to improve democratic control over both data and other bottlenecks such as compute. 

\begin{acks}
We benefited greatly from insightful comments from the following individuals: Lauro Langosco, Usman Anwar, Shahar Avin, Stella Biderman, Henry Ashton, Micah Carroll, Yawen Duan, David Krueger, Robert Harling.
\end{acks}

\bibliographystyle{ACM-Reference-Format}
\bibliography{ARXIV}

\appendix

\end{document}